\documentclass[pra,twocolumn,showpacs,preprintnumbers,amsmath,amssymb]{revtex4}

\usepackage{graphicx}
\usepackage{dcolumn}
\usepackage{bm}

\newcommand{\Tr}{\mathop{\text{Tr}}\nolimits}
\newcommand{\mapright}[1]{\smash{\mathop{\hbox to 1.0cm{\rightarrowfill}}\limits^{#1}}}

\begin{document}


\title{Information Theoretical Approach to  Control of Quantum Mechanical Systems}

\author{Shiro Kawabata}
 \email{s-kawabata@aist.go.jp}
\affiliation{%
Nanotechnology Research Institute (NRI) and Research Consortium for 
Synthetic Nano-Function Materials Project (SYNAF), National Institute of 
Advanced Industrial Science and Technology (AIST), \\1-1-1 Umezono, Tsukuba, 
Ibaraki 305-8568, Japan 
}%

\date{\today}

\begin{abstract}
Fundamental limits on the controllability of quantum mechanical systems are discussed in the light of quantum information theory.
It is shown that the amount of entropy-reduction that can be extracted from a quantum system by feedback controller is upper bounded by a sum of the decrease of entropy achievable in open-loop control and the mutual information between the quantum system and the controller.
This upper bound sets a fundamental limit on the performance of any quantum controllers whose designs are based on the possibilities to attain low entropy states.
 An application of this approach pertaining to quantum error correction is also discussed. 
\end{abstract}

\maketitle

%
%
%
\section{Introduction}
%
%
%
Ever since the discovery of quantum mechanics, the problem of controlling quantum systems has been an
important experimental issue~\cite{rf:quantumcontrol1,rf:Rabitz}.
A variety of techniques are available for controlling quantum
systems, and a detailed theory of quantum control has been developed~\cite{rf:quantumcontrol2,rf:rice,rf:LANL,rf:shapiro}.
Methods of geometric and coherent control are
particularly powerful for controlling the coherent dynamics of quantum
systems~\cite{rf:Rabitz,rf:Lloyd1}.

The rapid development of quantum information technology suggests that quantum control theory might profitably be reexamined from the perspective of quantum information theory~\cite{rf:QI,rf:NC}.
In this paper we address explicitly the role of quantum information and entropy in quantum control processes.
Specifically, based on classical theories~\cite{rf:Weidemann,rf:Poparavskii,rf:Lloyd2,rf:Touchette1, rf:Touchette2}, we prove several limiting results relating to the ability of a control device to
reduce the von Neumann entropy $S=\Tr \rho^Q \log \rho^Q $ of an arbitrary quantum system $\rho^Q$ in the cases where (i) a
controller independently acts to the state of the system (open-loop
control) and (ii) the control action is influenced by some information
gathered from the system (feedback control).

When a quantum system $Q$ initially prepared in a pure state $\rho_0$ interacts with an environment represented by the density operator $\rho^E$, the system $Q$ and environment evolve according to  the joint unitary evolution operator $U_{QE}$.
Then the density operator for the system $Q$ and environment  is
\begin{equation}
\rho = U_{QE} \left( \rho_{0} \otimes \rho^E \right)  U_{QE}^\dagger
.
\end{equation}
After performing a partial trace over environment variables, the marginal density matrix of the system $Q$ is represented  by a completely positive and trace preserving map ${\cal E}$,
which takes the form
\begin{equation}
\rho^Q = {\cal E} \left( \rho_{0} \right)   = \sum_{i} E_i  \rho_{0} E_i^\dagger
,
\end{equation}
where the Kraus operators $E_i$'s satisfy the trace preserving property, $i.e.,$ $\sum_i E_i^\dagger E_i=I$.
This equation is known as operator-sum representation of the quantum operation ${\cal E}$.
Unitary evolution of the quantum system is a special case in which there is only one non-zero term in the operator sum.
On the other hand, if there are two or more terms, the pure initial state becomes a mixed state.
Therefore, the von Neumann entropy of the system $Q$ increases, $i.e.$, $S(Q) \equiv S(\rho^Q) > S(\rho_0)$, because of the interaction with environment.
In this paper, we define the purpose of quantum control as a reduction of the entropy of the
system $Q$~\cite{rf:Touchette1, rf:Touchette2}, $e.g.$, quantum Maxwell demon~\cite{rf:maxwell1,rf:maxwell2}, quantum bang-bang control~\cite{rf:bang1,rf:bang2}  and quantum error correcting code~\cite{rf:qec1,rf:qec2}.
In the following, we shall show the information-theoretic analysis of open-loop and closed-loop (feedback) control, and give the fundamental limits on the control of quantum mechanical systems from the viewpoint of quantum information theory.

%
%
%
\section{Quantum Open-Loop Control}
%
%
%

Here, we present information-theoretic analysis of the quantum open-loop control.
First, we shall look at a joint unitary evolution (a control unitary operation) of the quantum system $Q$ and controller $C$.
Let  the quantum system $Q$ and the controller $C$  be disentangled before the control unitary operation.
We also assume that the states of system $Q$ and $C$ are respectively given by eq.~(2) and
\begin{equation}
\rho^C = \sum_i p_i  | i \rangle_C \langle i |
.
\end{equation}
Here $ | i \rangle_C$ is an orthonormal basis of system $C$ and $\sum_i p_i=1$.
Theretofore, the state of the joint system $QC$ is  given by
\begin{equation}
\rho^{QC}=\rho^{Q} \otimes \rho^{C}
=\sum_{i,j} p_i \rho_j^Q \otimes  | i \rangle_C \langle i |
,
\end{equation}
where $ \rho_j^Q \equiv E_j \rho_0 E_j^\dagger$.
In order to reduce the entropy of the system $Q$, a control unitary transformation $U_\text{open}$ is applied to joint system  $QC$.
Then, the system  $QC$ undergoes the evolution:
\begin{equation}
\rho^{QC} \to U_\text{open} \rho^{QC} U_\text{open}^\dagger
.
\end{equation}
 In the following we shall consider two types of control unitary operation, $i.e.,$ global unitary operation (Fig.~1) and LOCC (local quantum operation and classical communication) (see Fig.~2).
 In the former case, the entropy of the total system becomes
\begin{equation}
S(Q,C)
=
S(Q_\text{out},C_\text{out})
\le
S(Q_\text{out})+S(C_\text{out}),
\end{equation}
where we have used the subadditivity of the entropy.
From this inequality, we finally obtain the entropy reduction as
\begin{equation}
\Delta S_Q^\text{open}
\equiv
S(Q)-S(Q_\text{out})
\le
S(C_\text{out})-S(C)
,
\end{equation}
with equality if and only if  $\rho^{Q_\text{out}C_\text{out}}=\rho^{Q_\text{out}} \otimes \rho^{C_\text{out}}$.
Therefore, the entropy reduction is upper bounded by the maximum amount of the entropy increase of $C$.

 On  the other hand, in the case of LOCC strategy (Fig.~2), the control unitary operation is given by
\begin{equation}
U_\text{open} = \sum_i U_i \otimes  | i \rangle_C \langle i |
.
\end{equation}
Therefore, the state after the open loop control becomes
%
%
%
%
%
%
%
%
\begin{figure}[t]
\begin{center}
\includegraphics[width=6cm]{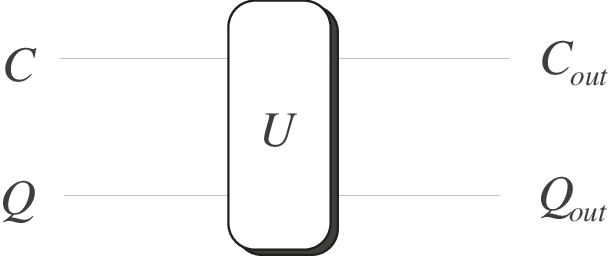}
\end{center}
\caption{Quantum open-loop control using global unitary transformation.}
\label{f1}
\end{figure}
%
%
%
%
%
\begin{figure}[t]
\begin{center}
\includegraphics[width=6cm]{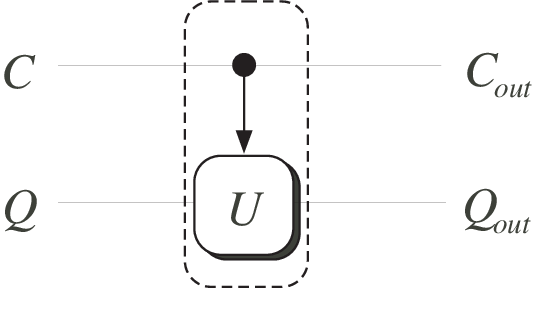}
\end{center}
\caption{Quantum open-loop control using LOCC.}
\label{f2}
\end{figure}
%
%
%
%
\begin{equation}
\rho^{Q_\text{out}C_\text{out}}
=
\sum_{i,j} p_i U_i \rho_j^Q  U_i^\dagger \otimes | i \rangle_C \langle i |
.
\end{equation}
Then, the marginal density operator of $Q_\text{out}$ is given by
\begin{equation}
\rho^{Q_\text{out}}
=
\Tr_{C} \rho^{C_\text{out}Q_\text{out}}
=
\sum_i p_i U_i \rho^Q U_i^\dagger
.
\end{equation}
Now using the  concativity of the von Neumann entropy $S \left( \sum_i p_i  \rho_i \right) \ge  \sum_i p_i S (\rho_i)$, we see that
\begin{align}
S(Q_\text{out})
&=
S \left( \sum_i p_i U_i \rho^Q U_i^\dagger \right)
\nonumber\\
&\ge
\sum_i p_i S \left(  U_i \rho^Q U_i^\dagger \right)
=
S \left(Q \right)
.
\end{align}
Therefore for open-loop control using the LOCC strategy, we finally obtain
\begin{equation}
\Delta S_Q^\text{open} \le 0
.
\end{equation}
This means that we can never reduce the  entropy of system $Q$ in contrast with the case of the global unitary  operation strategy.

%
%
%
\section{Quantum Feedback Control}
%
%
%

Next we shall consider the quantum feedback control.
In this case, the controller $C$  performs measurements on the system $Q$ and feeds back the results of these measurements by applying operations that are the functions of the measurement results.
Although both the system $Q$ and the controller $C$  are quantum mechanical in principle, the feedback operations we consider here involve feeding back classical information.
The feeding back quantum information via fully coherent quantum feedback was recently discussed by Lloyd and co-workers~\cite{rf:Lloyd1,rf:LS} and experimentally demonstrated~\cite{rf:Nelson}.

To analyze quantum feedback control, we need to consider quantum measurement processes.
For simplicity, we consider a positive operator-valued measure measurement in which the entropy of the system $Q$ dose not decreases, $e.g.$, the conventional von Neumann measurement.

As in the case of preceding section, we shall investigate two types of control strategies (Figs.~3 and 4).
A basic quantum feedback control using a global control unitary operation is presented in Fig.~3.
Then the entropy of $C_\text{out}$ is calculated as
\begin{align}
S(C_\text{out})
={}&
S(Q_\text{out},C_\text{out}) - S(Q_\text{out})
+ I(Q_\text{out}:C_\text{out})
\nonumber\\
={}&
S(Q',C') - S(Q_\text{out}) + I(Q_\text{out}:C_\text{out})
\nonumber\\
\le{}&
S(Q) -S(Q_\text{out}) +S(C') - I(Q':C')
\nonumber\\
&{}+ I(Q_\text{out}:C_\text{out})
,
\end{align}
where $I(A:B)=S(A)+S(B)-S(A,B)$ is the quantum mutual information of systems $A$ and $B$~\cite{rf:NC}.
Therefore, the entropy reduction for quantum feedback using the global unitary operation is given by
\begin{align}
\Delta S_Q^\text{feedback}
={}&
S(Q) -S(Q_\text{out})
\nonumber\\
\le{}&
S(Q',C') - S(Q_\text{out})
 + I(Q_\text{out}:C_\text{out})
\nonumber\\
={}&
S(C_\text{out})-S(C') -  I(Q_\text{out}:C_\text{out})
\nonumber\\
&{}+  I(Q':C')
\nonumber\\
\le{}&
\max_U \Delta S_Q^\text{open} +  I(Q':C')
.
\end{align}
Here $\max_U \Delta S_Q^\text{open}$ is the maximum entropy reduction attained by restricting the control model to  open-loop system.
The equality holds if and only if $\rho^Q=\rho^{Q'}$ and $S(C_\text{out})-S(C') -  I(Q_\text{out}:C_\text{out}) = \max_{U} \Delta S_Q^\text{open}$.
Therefore, the maximum improvement that closed-loop can give over open-loop control is limited by the quantum mutual information obtained by the controller $C$.

%
%
%
\begin{figure}[t]
\begin{center}
\includegraphics[width=6cm]{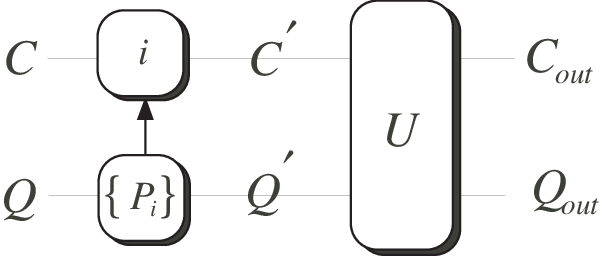}
\end{center}
\caption{Quantum feedback control using global unitary transformation.}
\label{f3}
\end{figure}
%
%
%
%
\begin{figure}[t]
\begin{center}
\includegraphics[width=6cm]{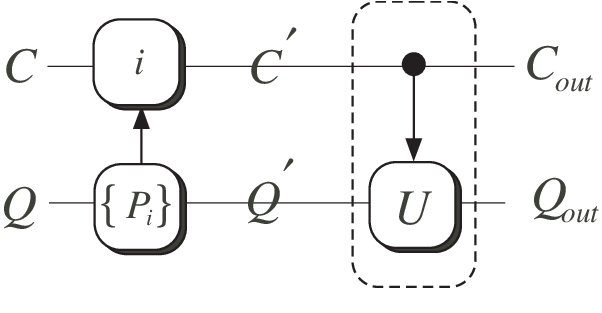}
\end{center}
\caption{Quantum feedback control using LOCC.}
\label{f4}
\end{figure}
%
%

Next we shall consider quantum feedback control using the LOCC (see Fig.~4),
and show that the entropy reduction is upper bounded by the quantum mutual information between $Q'$ and $C'$, $i.e.$, $\Delta S_Q^\text{feedback} \le I(Q':C')$.

In this strategy, one performs a measurement (on the state $\rho^Q$) described by positive operators  $\{P_i\}$, and feeds back the results by applying a unitary transformation $U_i$ when the $i$th outcome is found.
Then the state change of the subsystem $Q$ can be written as
\begin{eqnarray}
      \rho^{Q}
      &\to&
      \rho^{Q'} =\sum_i   P_i \rho^Q P_i^\dagger
      \\
       &\to&
      \rho^{Q_\text{out}} = \sum_i U_i  P_i \rho^Q P_i^\dagger U_i ^\dagger
      \equiv {\cal C} (\rho^Q).
\end{eqnarray}
From the inequality of the entropy exchange $S_e(\rho, {\cal E})$ for a quantum operation ${\cal E}$~\cite{rf:Nielsen},
\begin{equation}
S \left(  {\cal E} (\rho) \right) - S(\rho) + S_e(\rho, {\cal E} ) \ge 0,
\end{equation}
it follow that
\begin{equation}
S \left(  Q_\text{out}  \right) - S(Q) + S_e(\rho^Q, {\cal C} ) \ge 0
.
\end{equation}
Thus we have inequality for the entropy reduction,
\begin{equation}
\Delta S_Q^\text{feedback} = S(Q)- S(Q_\text{out}) \le S_e(\rho^Q, {\cal C} )
.
\end{equation}
The entropy exchange is not greater than the Shannon entropy for the probabilities $r_i = \Tr (U_i P_i \rho^Q P_i^\dagger U_i^\dagger)$~\cite{rf:Nielsen}.
Thus,
\begin{equation}
S_e(\rho^Q , {\cal C} )  \le H(r_i)
,
\end{equation}
where equality holds  if and only if the operator $U_i P_i$ are a canonical decomposition of ${\cal C}$ with respect to $\rho^Q$~\cite{rf:Nielsen}.
Therefore we obtain
\begin{equation}
\Delta S_Q^\text{feedback}  \le H(r_i)  = - \sum_i r_i \log r_i
.
\end{equation}
We can also show  the equality $H(r_i) = I(Q':C')$~\cite{rf:proof}. 
That is, in the case of the quantum feedback using LOCC, the entropy reduction is given by
\begin{equation}
\Delta S_Q^\text{feedback} \le   I(Q':C')=H(r_i)
.
\end{equation}
This implies  that  the maximum amount of entropy reduction is exactly equal to  the quantum mutual information between subsystems $Q'$ and $C'$, $i.e.$, $I(Q':C')$.

The quantum mutual information  $I(A:B)$  is related to correlation between subsystems  $A$ and $B$~\cite{rf:NC}.
If a joint system $A$ and $B$ is a product state, then  $I(A:B)=0$.
However, $I(A:B)>0$ if the subsystems $A$ and $B$ are  (classically or quantum mechanically) correlated.
In the case of the quantum feedback shown in Figs.~3 and 4, the quantum measurement germinates not quantum but classical correlation between $Q$ and $C$.
Therefore, we can conclude that the classical correlation between $Q'$ and $C'$ can increase the amount of the entropy reduction in compared with the case of the quantum open loop control.

Finally, we shall show the information theoretical analysis of quantum error correction~\cite{rf:NC,rf:QEC} as an example of the quantum feedback control using LOCC (Fig. 4).
The quantum error correction can be thought of as a type of refrigeration process, capable of keeping a quantum system $Q$ at a constant entropy, despite the influence of noise processes which tend to increase the entropy of system $Q$.
General error correction procedure is described by following three steps ((a) error (b) syndrome measurement and (c) error correction)~\cite{rf:Ekert,rf:Vedral},
\begin{eqnarray}
&&
\rho_0 = | \psi \rangle_Q \langle \psi| \otimes  | 0 \rangle_M \langle 0 | 
	  \nonumber\\
&\stackrel{(a)}{\longrightarrow}&
      \rho = \sum_i p_i e_i | \psi \rangle_Q \langle \psi| e_i^\dagger  \otimes  | 0 \rangle_M \langle 0 | 
	  \nonumber\\
&\stackrel{(b)}{\longrightarrow}&
      \rho' = \sum_i p_i e_i | \psi \rangle_Q \langle \psi| e_i^\dagger  \otimes  | i \rangle_M \langle i | 
	  \nonumber\\
&\stackrel{(c)}{\longrightarrow}&
      \rho_{out} =  | \psi \rangle_Q \langle \psi|  \otimes \sum_i p_i  | i \rangle_M \langle i |
	  ,
\end{eqnarray}
where $| \psi \rangle_Q$ is the (initial) encoded quantum state, $e_i $ is the unitary error operator which acts on the state of system $Q$ only and $ | i \rangle_M$ is the orthonormal basis of a measurement apparatus which acts as the controller $C$.
In the language of the control theory, steps (b) and (c) correspond to the feedback and the control operation process, respectively.
In the following, we shall show an entropic analysis of the quantum error correction and show that equality of eq.(22) holds.
For simplicity, we restrict our analysis to the case of non-degenerate quantum code.
From eq.(23), we obtain the reduced density operator of the quantum system $Q$ as 
\begin{eqnarray}
\rho^Q &=& \Tr_M \rho = \sum_i p_i e_i | \psi \rangle_Q \langle \psi| e_i^\dagger,\\
\rho^{Q_{out}} &=& \Tr_M \rho_{out} =   | \psi \rangle_Q \langle \psi| 
.
\end{eqnarray}
Therefore, the entropy reduction is given by
\begin{eqnarray}
\Delta S_Q^{QEC} & \equiv &S (Q)-S(Q_{out}) 
\nonumber\\
&=&
H(p_i) +  \sum_i p_i S(\rho^Q_i)
=
H(p_i)
,
\end{eqnarray}
where $\rho^Q_i \equiv e_i | \psi \rangle_Q \langle \psi | e_i^\dagger$.
On the other hand, quantum mutual information before step (c) is $I(Q':C') =  H(p_i)$.
Therefore we have
\begin{eqnarray}
 \Delta S_Q^{QEC} =H(p_i)=I(Q':C')
 ,
\end{eqnarray}
which is the desired result, i.e., equality of eq.(22) holds.
This means that the quantum error correction is an optimal quantum feedback control system from the view points of the information theory.

%
%
%
\section{Summary}
%
%
%

In summary, we have analyzed the quantum control system by use of the quantum information theory and showed information theoretical limits of control of quantum mechanical systems.
By applying our approach to an entropic analysis of quantum error correcting code, we have showed that we can regard  the quantum error correction procedure as an  (information  theoretically) optimal quantum feedback control system.
Our result will also help in understanding quantum bang-bang control and quantum Maxwell demon.

I would like to thank S. Abe for useful discussion.
This work was supported by NEDO under the Nanotechnology Materials Program.


%
%


\begin{thebibliography}{99}
\bibitem{rf:proof}
In order to obtain the equation $H(r_i) = I(Q':C') $, we notice that the joint state after measurement is given by
$
\rho^{Q' C'}
=
\sum_i P_i \rho^Q P_i^\dagger \otimes | i \rangle_C \langle i |
=
\sum_i r_i \rho_i \otimes | i \rangle_C \langle i |
,
$
%
%
%
%
where $\rho_i \equiv P_i \rho^Q P_i^\dagger/r_i$.
Therefore the entropy of the system $Q'C'$ is given by
%
%
%
$ 
S(Q',C')
=
S (\sum_i r_i \rho_i \otimes | i \rangle_C \langle i |  )
=
H(r_i) + \sum_i r_i S (\rho_i)
=
H(r_i)
,
$ 
%
%
%
%
where in the second equality we have used the joint entropy theorem~\cite{rf:NC}.
The marginal density operators of subsystems $Q'$ and $C'$ are
respectively given by $
\rho^{Q'}
=
\sum_i r_i  | i \rangle_Q \langle i |
$
and
$
\rho^{C'}
=
\sum_i r_i  | i \rangle_C \langle i |
$
,
%
%
%
%
so we have
%
%
%
$S(Q')=S(C')=H(r_i)$.
%
%
%
%
Therefore the mutual information between subsystems $Q'$ and $C'$ is $I(Q':C')= S(Q') + S(C')-S(Q',C')=H(r_i)$.
\end{thebibliography}
\end{document}